# Band restructuring of ordered/disordered blue TiO$_2$ for visible photocatalyst

Simgeon Oh[a,c,#], Ji-Hee Kim[a,c,#,*], Hee Min Hwang[a,c], Doyoung Kim[a,c], Joosung Kim[a,c], G. Hwan Park[a,b], Joon Soo Kim[a,c], Young Hee Lee[a,c,*], Hyoyoung Lee[a,b,*]

[a]*Center for Integrated Nanostructure Physics (CINAP), Institute for Basic Science (IBS), Sungkyunkwan University, Suwon 16419, Korea*
[b]*Department of Chemistry, Sungkyunkwan University, Suwon 16419, Korea*
[c]*Department of Energy Science, Sungkyunkwan University, Suwon 16419, Korea*

[#]These authors contribute equally to this work.
***Corresponding authors:** kimj@skku.edu (Prof. J.-H. Kim)

leeyoung@skku.edu (Prof. Y. H. Lee)

hyoyoung@skku.edu (Prof. H. Lee)

**Abstract**

Black TiO$_2$ with/without noble metal has been proposed for visible photocatalyst, still leaving poor catalyst efficiency. Alternatively, phase-mixed TiO$_2$ such as anatase and rutile has been commonly used for visible catalysts with the inevitable inclusion of noble metal. Here, we perform a noble metal-free visible photocatalyst blue TiO$_2$ with type-II band-aligned ordered anatase/disordered rutile structure, via phase-selective reduction with alkali metals. The changed band alignment in this heterostructure was identified by absorption and ultraviolet photoemission spectroscopy, which was further confirmed by transient charge separation. The band alignment of type-I and type-II was clearly restructured by converting from ordered to disordered phase with a prolonged reduction period and as followed light absorbance enhancement also observed. Initiated type-I in a pristine sample, the type-II was organized from disordered rutile phase in 3-day Li-reduction. The type-II disordered rutile TiO$_2$ heterostructure exhibits a remarkable photocatalytic performance by 55 times higher than conventional P25 $TiO_2$ in solar-light driven hydrogen evolution reaction owing to an efficient electron and hole separation of type-II heterojunction. Furthermore, this restructured heterojunction type-II $TiO_2$ demanded 10 times less Pt amount as a co-catalyst for the comparable photocatalytic performance, compared to Pt decorated type-I pristine anatase/rutile phase-mixed $TiO_2$.

**Keywords**: Phase-selective reduction, Phase-mixed $TiO_2$, Energy band restructuring, Photocatalysis, Hydrogen evolution reaction

## 1. Introduction

While titanium dioxide ($TiO_2$) has been widely used as a photocatalyst under UV light, its photocatalytic efficiency is poor even in the presence of a noble metal [1-3]. The visible light solar spectrum is an ideal option to improve the photocatalyst efficiency by ~40% resulting from the solar irradiation intensity compared to 4-5% in UV spectra. For the visible light photocatalyst, a single phased black $TiO_2$ has been explored with/without noble metals for over a half-century, but required realistic efficiency has not been obtained [4,5]. Afterward, to make a black $TiO_2$ through whole surface modification of two phase-mixed $TiO_2$ that is one of renowned photocatalyst material with large surface area, high reduction potential, and efficient charge transfer property from different phase junction structures has been also investigated to improve photocatalytic performance [6,7]. However, this black $TiO_2$ shows the core-shell structure as a result of the non-phase-selective surface reduction, exposing only the disordered $TiO_2$ surface. Therefore, its photocatalytic efficiency is still poor mainly due to an easy charge recombination on the disordered shell surface [8]. To overcome the charge recombination issue, two different facet structures of phase-selectively disordered anatase/rutile $TiO_2$ are previously reported, showing a superior photocatalytic performance than the core-shell structure [9,10].

Nevertheless, in the phase-selectively disordered anatase/rutile $TiO_2$ facet structure, the relationship between type-I and type-II of the restructured heterojunction band structure and charge transfer efficiency with prolonged reduction time is not clearly understood. For example, the type-II band alignment in the pristine anatase/rutile $TiO_2$ heterostructure has been proposed for the efficient extraction of electrons and holes [11-13]. In contrast, type-I band alignment has been suggested in photocatalyst systems due to low-energy electron transfer from the rutile to anatase phase [14-16]. To make a worse, for the phase-selectively disordered anatase/rutile $TiO_2$, there is no report about the relationship between the band types and charge transfer efficiency. Thus, two questions remain to be resolved: i) which type of band alignment, type-I or type-II, of the anatase/rutile junction and phase-selectively disordered in $TiO_2$? and ii) which phase, anatase or rutile, plays the dominant role for enhancing the charge transfer efficiency through a proper band restructuring?

Here, we report the heterojunction energy band restructuring of selectively ordered/disordered anatase/rutile $TiO_2$ by tailoring phase-selective reduction. We find obvious type-II heterojunction band structure is organized by 3-day Li-reduction, that the restructured type-II heterojunction is the best choice for the photocatalyst with fast charge transfer between disordered rutile and ordered anatase phase. We were able to tailor the HOMO and LUMO positions by XPS and UPS measurements. Carrier population dynamics clearly confirm by femtosecond transient absorption with the time evolution of band alignment for type-I and type-II. The type-II disordered rutile/ordered anatase $TiO_2$ heterostructure further demonstrates the best photocatalytic performance in the visible range to date.

**2. Experimental**

*2.1. Chemicals and Materials*

The anatase/rutile phase-mixed $TiO_2$ (Degussa P25) powder was purchased from the Degussa Company. Anatase phase $TiO_2$, rutile phase $TiO_2$ powder, lithium metal, sodium metal, chloroplatinic acid, sodium sulfate, and hydrochloric acid were purchased from Merck, Inc. Ethylenediamine was procured from Tokyo Chemical Industry. All reagents were used without further purification.

*2.2 Material characterization*

The crystalline structures of the samples were analyzed by an X-ray diffraction diffractometer (XRD, Rigaku - Ultima IV) with Cu Kα radiation (λ = 0.15418 nm) at room temperature. All the core-level binding energies were measured by X-ray photoelectron spectroscopy (XPS, VG Microtech - ESCA 2000) with Mg Kα (= 1253.84 eV) radiation referenced to the C 1s peak of surface adventitious carbon at 284.6 eV. Survey scans and detailed scans of the Ti 2p, O 1s photoelectron peaks were recorded for the samples. In addition, the valence energy band edge of the samples was evaluated by XPS valence band analysis. The optical absorption by the samples was characterized by diffuse reflectance spectroscopy (DRS, Shimadzu-UV-3600) with PTFE used as the white diffuse reflectance standard. The energy band gap of the catalysts was calculated from diffuse reflectance spectra in a Tauc plot. For transient absorption spectroscopy (TAS), the beam from a Ti:sapphire amplifier (Libra, Coherent, USA) with a 1-kHz repetition rate and 80 fs pulse duration was divided into two beams. Here, 95 % of the amplifier drives an optical parametric amplifier (TOPAS Prime, Light Conversion, Lithuania) and is used as a pump beam, which is tuned to 300 nm to excite the $TiO_2$ sample at an excitation intensity of 1 μJ/pulse. The rest of the amplifier beam produces a white light continuum probe beam in the visible spectral region. Transmission electron microscopy (TEM) was performed using a JEOL-JEM-2100F field-emission instrument at an acceleration voltage of 300 kV. Ultraviolet photoelectron spectroscopy (UPS, Thermo Fisher - ESCALAB 250 Xi) measurement was conducted with UV photons (21.22 eV for He I radiation). To separate the secondary edges of the samples, a negative potential of 5.0 V was applied for the optimization of Au spectrum. The work function of 5.1 eV was obtained which is a good agreement with previous reports. AC impedance data of the $TiO_2$ samples were collected using electrochemical impedance spectroscopy (EIS, CH Instruments - CHI660C) with a 0.5 M $Na_2SO_4$ solution in three-electrode mode with a 3.0 M Ag/AgCl reference electrode. A sinusoidal AC perturbation with a 50 mV amplitude was applied in the frequency range of 10 mHz to 100 kHz. The platinum weight ratio was confirmed by XPS, and the best performing Pt-deposited $TiO_2$ samples were measured by inductively coupled plasma optical emission spectrometry (ICP-OES, Agilent - 5100).

*2.3 Phase-selectively reduced $TiO_2$ preparation*

Phase-selectively reduced $TiO_2$ was prepared via a modified Birch reduction reaction with different alkali metal-ethylenediamine (EDA) solutions. Single-phase anatase or rutile $TiO_2$ was used for the phase-selective reduction reaction study, and Degussa P25 served as the standard anatase/rutile phase-mixed $TiO_2$ for investigation of the phase-selective band restructuring and its photocatalytic performance. Li-EDA and Na-EDA solutions were prepared as follows. Alkali metal (250 mg) was added to a 250 ml round bottom flask. The flask was well sealed with a rubber septum and purged with argon gas for 60 minutes. Subsequently, 50 ml of EDA was injected into the flask with a 50 ml syringe. This solution was vigorously stirred for 24 hours under an argon atmosphere to dissolve the alkali metal. Finally, 500 mg of the respective anatase, rutile, or phase-mixed $TiO_2$ powder was poured into the prepared solution and stirred for 1-7 days to reduce the $TiO_2$ crystalline phase. The Li-EDA solution was only used for rutile phase-selective reduction (Li-reduction), and the Na-EDA solution was only used for anatase phase-selective reduction (Na-reduction). These processes are referred to as $TiO_2$ phase-selective reduction with Li-EDA or Na-EDA solution treatment. The resulting product solution was transferred into a beaker and titrated with 1N HCl aqueous solution until the solution reached a pH of 7. The final blue $TiO_2$ powder product was filtrated with a membrane filter, washed with ethanol and deionized water, and dried in a vacuum oven at 40 °C.

*2.4 Pt photodeposition*

Platinum (Pt)-deposited $TiO_2$ samples were prepared by photodeposition. Pristine P25 $TiO_2$ or 3-day Li-reduced phase-mixed $TiO_2$ was used as the photocatalyst powder, and a 0.001 to 1.0 mM aqueous $H_2PtCl_6$ solution was added to the powder to deposit different amounts of Pt co-catalyst. $TiO_2$ powder (100 mg) was added to 100 ml of different concentrations of the Pt precursor solution in a 100 ml glass vial. $TiO_2$ was well-dispersed by sonication treatment for one hour, and 1 ml of methanol was added as a hole scavenger. The solution-containing vial was well sealed with a quartz cover, and the solution was irradiated with a 100 W UV light (Power Arc UV-100, UV Process Supply, Inc.) under vigorous stirring for one hour. The resulting Pt-deposited $TiO_2$ sample was filtered with a membrane filter and then washed with ethanol and deionized water.

*2.5 Photocatalytic performance estimate*

The photocatalytic hydrogen evolution reaction was conducted using a 100 ml Pyrex flask, and AM 1.5 solar light (Oriel Sol 3A Class AAA Solar Simulator, Newport Corp.) was used for the photocatalytic reaction. The newly prepared $TiO_2$ photocatalyst was soaked in 40 ml of a 3:1 v/v% deionized water/ethanol mixture and sonicated for 24 hours. Then, argon gas was bubbled out for one hour. Simulated 1 sun solar light was irradiated on the prepared samples to generate $H_2$ gas. The produced $H_2$ gas was sampled using a 1 ml gas-tight syringe and then analyzed by gas chromatography (YL6500GC, Young Lin Instrument Co. with a molecular sieve 13X column and TCD detector).

**3. Results and discussion**

## 3.1. Phase-selective reduction of phase-mixed $TiO_2$

Our phase-selective reduction of anatase/rutile phase-mixed $TiO_2$ nanomaterials and their energy band alignment were carefully examined. The reduced phase-mixed $TiO_2$ samples for phase-selective energy band restructuring are denoted as ordered anatase/disordered rutile ($A_o/R_d$) and disordered anatase/ordered rutile ($A_d/R_o$) [17,18]. Degussa P25 served as a standard anatase/rutile phase-mixed $TiO_2$ in this work. The heterojunction band structure of $TiO_2$ was considered from the respective band structure of the single-phase disordered (or ordered) anatase and rutile phase $TiO_2$ (Table S1, Fig. S1, S2, and S3). Representative heterojunction band structures of pristine phase-mixed $TiO_2$ and reduced $TiO_2$ are schematically shown in Fig. 1 including type-I with a straddling bandgap for centered ordered rutile and ordered anatase, obvious type-II with a staggered bandgap for disordered rutile and ordered anatase, and reverse type-I for ordered rutile and disordered anatase. The phase-mixed $TiO_2$ was reduced with Li and Na metals for 7 days. From the carrier dynamics point of view, a type-II heterostructure is ideal for electron transfer from disordered rutile to ordered anatase in the LUMO position and hole transfer back to disordered rutile from ordered anatase in the HOMO position. This is in contrast with a type-I heterostructure, where the transferred electrons and holes recombine with each other, suppressing catalyst efficiency.

The color of the reduced $TiO_2$ was converted to blue from white after reduction (Fig. S4). The color change is distinct from the genuine black color obtained from conventional hydrogen-reduced $TiO_2$ since the reduced two phase-mixed $TiO_2$ shows blue owing to one of the $TiO_2$ phases kept crystalline structure [19]. The XRD and HRTEM data in Fig. 2 show that an ordered (crystalline) anatase phase is identified with a lattice spacing of 3.53 Å along the (101) direction with a disordered (amorphous) rutile phase after 7 days of Li-reduction (Fig. 2b). Meanwhile, an ordered rutile phase was observed with a lattice spacing of 3.26 Å along the (110) direction with disordered anatase after 7 days of Na-reduction (Fig. 2d). Additional electron-diffraction FFT pattern information also showed obvious lattice planes of anatase (101) and (110) but amorphous rutile structure FFT pattern with week anatase overlapped plane for Li-reduction in the inset of Fig. 1b. The FFT pattern of Na-reduction, rutile (110) plane only showed at the rutile particle (see the insets in Fig. 2d). Such structural time evolutions of the reduction were confirmed by XRD (Fig. 2a and 2c). Clear peaks from anatase (101) at 2 theta = 25.3 degree and rutile (110) at 2 theta = 27.5 degree are identified in the phase-mixed $TiO_2$ sample. The peak ratio of the anatase to rutile phase is large due to the different volumetric ratio of anatase and rutile (80:20) in standard phase-mixed $TiO_2$ (P25) [20,21]. The peak intensity ratio becomes larger with evolving Li-reduction time owing to the developed disordered rutile phase, which is similarly observed from the full-width-at-half-maximum (FWHM) of the peak (Table S2). In contrast, with Na-reduction, the peak intensity ratio gradually decreases in the anatase phase with evolving reduction time, which is evidence of the emerging disordered anatase phase, whereas the change of the peak intensity is negligible in the rutile phase. Such an emerging disordered anatase phase is reflected by the decreasing peak intensity ratio of $I_A/I_R$ and the FWHM ratio (insets in Fig. 2c and Table S2).

The time evolution of oxygen vacancies and rich $Ti^{3+}$ oxidation states from the disordered phase is confirmed by X-ray photoelectron spectroscopy (XPS) in Fig. 3. The peak position of Ti 2p centered at binding energies of 457.7, 458.3, 463.2, and 464.1 eV corresponds to the $Ti^{3+}2p_{3/2}$, $Ti^{4+}2p_{3/2}$, $Ti^{3+}2p_{1/2}$ and $Ti^{4+}2p_{1/2}$, respectively, which is a good agreement with the previous reports [22,23]. The $Ti^{4+}$ oxidation states in the Ti $2p_{3/2}$ peaks near 458.3 eV are dominant with a minor $Ti^{3+}$ contribution near 457.7 eV in the phase-mixed $TiO_2$. With Li-reduction, $Ti^{3+}$ oxidation states are distinctly developed after reduction for 7-day, confirming the emergence of a disordered rutile phase with consistent time evolution and the Ti $2p_{1/2}$ peak also shows same manner (Fig. 3a and inset). The XPS study of reduced single-phase rutile and anatase $TiO_2$ suggests that the $Ti^{3+}$ oxidation mainly obtained from the rutile phase while the anatase phase indicates no appreciable oxidation states, maintaining the ordered anatase phase (Fig. S5a and S5c). This trend is also congruent with the developed $Ti^{3+}$ oxidation state or $Ti_2O_3$ peak in the O 1s spectra. O 1s spectra in Fig. 3a show distinct four oxygen peaks at 529.3, 530.8, 531.8, and 533.2 eV which represents the $TiO_2$, oxygen vacancy ($Ti_2O_3$) and surface oxygen species (-OH, $H_2O$), respectively [24]. It shows that the oxygen vacancy $Ti_2O_3$ peak relatively increased as a function of prolonged Li-reduction time. In contrast, in Na-reduction, the $Ti^{3+}$ oxidation states emerge from the disordered anatase phase after 7-day reduction, which is consistent with the related $Ti_2O_3$ peak in the O 1s spectra (Fig. 3b and inset) [25-27]. A similar trend of oxidation state to phase-mixed $TiO_2$ was observed in single-phase rutile and anatase $TiO_2$ study (Fig. S5b and S5d). The apparent oxidation states of $Ti^{3+}$ in the disordered rutile phase with Li-reduction and in the disordered anatase phase with Na-reduction are essential to tailoring the type of band structure and further improving visible light photocatalyst efficiency.

*3.2. Identification of energy band restructuring*

To tailor the band-type of the ordered/disordered anatase/rutile $TiO_2$, we measured the direct bandgap from the Tauc plot via $\sim(\alpha h\nu)^{1/2}$ using UPS and the valence band edge from XPS based on the single-phase $TiO_2$ (Table S3 and Fig. S3). Fig. 4 exhibits the HOMO and LUMO gaps in terms of reduction time [28-30]. As Li-reduction time is prolonged, the initial HOMO-LUMO gap of 3.0 eV in the ordered rutile phase is narrowed further and the HOMO and LUMO positions are upshifted in the disordered rutile phase [31,32]. Meanwhile, they are nearly constant with a HOMO-LUMO gap of 3.2 eV in the ordered anatase phase, independent of reduction time. The type-I heterostructure is constructed prior to reduction and is converted to a type-II heterostructure after 3-day Li-reduction (Fig. 4a). The HOMO-LUMO gap of disordered rutile is 2.9 eV, which is beneficial for a visible photocatalyst with fast charge transfer. The band type restructuring in Li-reduction is markedly different from that in Na-reduction. Both HOMO and LUMO positions with an initial HOMO-LUMO gap of 3.2 eV are gradually up-shifted in disordered anatase phase, retaining the type-I heterostructure for up to 1-day with Na-reduction. The type-I is converted to type-II after 2-day reduction by switching the HOMO positions of disordered anatase and ordered rutile. Then, it is further transformed to reverse type-I after 6-day reduction, i.e., the LUMO position of disordered anatase is downshifted below that of ordered rutile (Fig. 4b). The HOMO-LUMO gap is further reduced to 2.7 eV in disordered anatase at 7-day Na-reduction, and band gap decrease was also observed in reduced $TiO_2$ (Fig. S6).

### 3.3. Charge extraction in heterostructure

Structural transformation and type-conversion of ordered/disordered anatase/rutile $TiO_2$ were identified with reduction time evolution, but a comprehensive understanding of the electron-hole separation process in each type of two phase-mixed $TiO_2$ remains elusive. To monitor the photoexcited charge extraction in each type, we performed femtosecond pump-probe spectroscopy and tracked the differential absorption changes ($\Delta A$) in the time domain [33]. The population density of the anatase phase was normalized to that of the rutile phase within the range of 1,000-2,000 ps near carrier recombination. The normalized kinetics of the anatase and rutile phase in phase-mixed $TiO_2$ with a type-I heterostructure was observed with a probe photon energy of 3.2 and 3.0 eV, respectively (Fig. 5a). The kinetics of the anatase phase (a probe photon energy of 3.2 eV) decays monotonically through a recombination channel with a maximum population near 0.3 ps [34]. Meanwhile, the maximum population density, $\Delta A_{max}$, is delayed to ~ 0.8 ps in the rutile phase. The carriers populated in the LUMO and HOMO levels of the anatase phase are transferred to those of the rutile phase, thereby enhancing the carrier density in the rutile phase. The charge transfer occurs efficiently at the interface, which is much shorter than the electron-hole recombination at the nanosecond time scale [35]. While the decay kinetics of each anatase phase is independent of Li-reduction time (Fig. 5b), the kinetic profiles of the rutile phase are dramatically changed. The population in the rutile phase is enhanced until the 1-day Li-reduction due to the electron and hole transfer (type-I). The population enhancement in the rutile phase becomes mundane from 3 to 7-day reduction. In general, charge separation at the interface of a type-II structure is greatly assisted by a large offset of band restructuring (Fig. 1, Left). However, the defective rutile formed at the interface in the disordered rutile/ordered anatase type-II structure can act as an electron-hole recombination center simultaneously, confirmed by the decreased population at 3 and 7-day reduction (blue dots in Fig. 5d). Moreover, the delayed population of both phases is prolonged when band alignment is restructured from type-I to type-II (Fig. 5b and blue dots in Fig. 5e). In contrast, the kinetic profiles of both phases with Na-reduction show quite different behaviors than Li-reduction (Fig. 5c). The population enhancement in either the anatase or rutile phase via charge transfer is indistinct over the Na-reduction time as well as the type-conversion (red open circles in Fig. 5d). The population density rapidly increases in the anatase phase from reverse type-I at 7-day Na-reduction (Fig. 4b and Fig. 5d). Simultaneously, the maximum population in the rutile phase is delayed by over 1 ps via slow carrier transfer to anatase from rutile (Fig. 5e), originating from disordered anatase with carrier localization centers (ex. oxygen vacancy), which was also confirmed by FTIR (Fig. S7).

*3.4. Photocatalytic hydrogen evolution*

We evaluated the photocatalytic hydrogen evolution reaction (p-HER) performance with the energy band restructured anatase/rutile phase-mixed $TiO_2$ as a proof of concept. The p-HER performance is primarily responsible for efficient charge separation which is achieved by proper band alignment and it is expected a type-II heterojunction [36-38]. The p-HER performance examined after confirming the reduction potential of each single-phase $TiO_2$ for hydrogen reduction by UPS measurement (Fig. S8 and S9). And then the evaluated p-HER performance of $TiO_2$ samples which are summarized in Fig. 6a and 6b. The ordered/disordered anatase/rutile $TiO_2$ reveals enhanced hydrogen production compared to pristine phase-mixed $TiO_2$. The disordered rutile phase with 3-day Li-reduction yields the highest p-HER performance (Fig. 6a). The hydrogen evolution rate under simulated 1 sun solar light is 707.03 $\mu mol\ g^{-1}\ h^{-1}$, which is 55 times higher than that of phase-mixed $TiO_2$ (12.72 $\mu mol\ g^{-1}\ h^{-1}$). This is mainly attributed to the enhanced light absorption (Fig. S6) and restructured type-II heterojunction structure between anatase and rutile after 3-day Li-reduction (Fig. 4a) for the efficient charge transfer (Fig. 5b). Meanwhile, the p-HER performance decreases rapidly right after 3-day Li-reduction. The degraded performance originates from a heavily defective amorphous structure, which is well comprehended by previous studies [39,40]. The rutile phase is further disordered with a longer Li-reduction time, thus increasing resistance, which is confirmed by the increased equivalent serial resistance determined from electrochemical impedance spectroscopy (Fig. S10a and Table S4) [41]. The p-HER rate with Na-reduction is similar to that of Li-reduction but significantly decreased with higher resistance (Fig. 6b). The large portion of the disordered anatase phase (80%) compared to the disordered rutile phase (20%) in the standard phase-mixed $TiO_2$ (P25) can be ascribed to the high resistance. Consequently, the p-HER performance decreased rapidly with prolonged Na-reduction time. The resistance becomes saturated after 5-day Na-reduction (Fig. 6b and Fig. S10b), and the LUMO level is reversed into a reverse type-I heterojunction to result in poor charge transfer (Fig. 4b). Rutile phase disordered $TiO_2$ shows the best p-HER performance at 3-day Li-reduction due to low resistance of the small disordered area, among others, while keeping type-II band alignment for efficient charge transfer (Fig. S11).

To investigate the tangible efficiency of photocatalyst, we explored the p-HER performance of 3-day Li-reduced phase-mixed $TiO_2$ with the smallest Pt co-catalyst amount. Pt has often served as a noble metal co-catalyst for HER active sites. The Pt-deposited pristine phase-mixed $TiO_2$ (Pt-$TiO_2$, type-I) without reduction shows tenable p-HER performance with an $H_2$ evolution rate of 5.714 mmol g$^{-1}$ h$^{-1}$ at a 0.03 mM Pt precursor solution. The weight ratio of Pt on the Pt-TiO$_2$ is 0.46 wt.% versus TiO$_2$, which is similar to previous reports (Fig. S12 and Table S5) [42-44]. In contrast, with a 10 times more dilute Pt precursor solution (0.003 mM, 0.043 wt.% of Pt), where the weight ratio of Pt to TiO$_2$ was analyzed by inductively-coupled-plasma atomic-emission spectroscopy, the p-HER rate of Pt-deposited ordered/disordered phase-mixed TiO$_2$ with 3-day Li-reduction (Pt-rTiO$_2$, type-II) shows the best performance (6.841 mmol g$^{-1}$ h$^{-1}$), which is greater than that of Pt-TiO$_2$ without reduction (Fig. 6c and Fig. S13).

## 4. Conclusions

In summary, we perform the band heterojunction restructuring between type-I and type-II in an ordered/disordered anatase/rutile TiO$_2$. The Li-reduction shows type-I and type-II, and the Na-reduction shows type-I, Type-II, and further reverse type-I heterojunction structure. The HOMO and LUMO levels are determined from the valence band edge using XPS and the bandgap from UPS analysis and restructured heterojunction band structures are confirmed by femtosecond transient absorption spectroscopy. We demonstrated that the type-II rutile phase disordered TiO$_2$ with Li-reduction is the best candidate for efficient charge transfer and minimum resistance with enhanced light absorption, yielding an HER rate 55 times higher than that of pristine phase-mixed TiO$_2$. Furthermore, the best photo-HER rate was realized as 6.841 mmol g$^{-1}$ h$^{-1}$ with 10 times smaller Pt amount compared to that in pristine TiO$_2$ with Pt. This provides plenty of room for phase-selective energy band restructuring to guide the ultimate design of photocatalytic materials with not only a TiO$_2$ heterostructure but also transition metal chalcogenides and perovskites.

**Credit Author Statement**

**Simgeon Oh** carried out all the experimental works. **Hee Min Hwang** and **Joosung Kim** synthesized catalysts. **Hee Min Hwang** performed DRS and band modeling. **Doyoung Kim** performed electrochemical experiments. **G. Hwan Park** performed XRD analysis. **Ji-Hee Kim** and **Joon Soo Kim** carried out the fs-TAS and provide interpretation. **Young Hee Lee** and **Hyoyoung Lee** contributed materials and analysis tools. **Simgeon Oh**, **Ji-Hee Kim**, **Young Hee Lee** and **Hyoyoung Lee** wrote the paper.

**Credit Author Statement**

The authors declare that they have no known competing financial interests or personal relationships that could have appeared to influence the work reported in this paper.

**Acknowledgements**

This work was supported by the Institute for Basic Science (IBS-R011-D1).

**Figures**

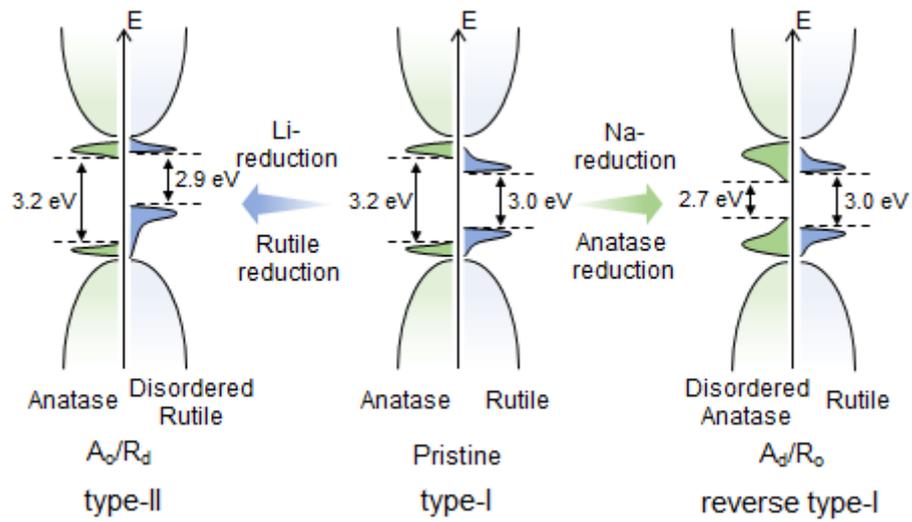

**Fig. 1.** Heterojunction band alignments of pristine phase-mixed $TiO_2$, band restructured 7-day Li-reduction ($A_o/R_d$, Left), 7-day Na-reduction ($A_d/R_o$, Right), respectively.

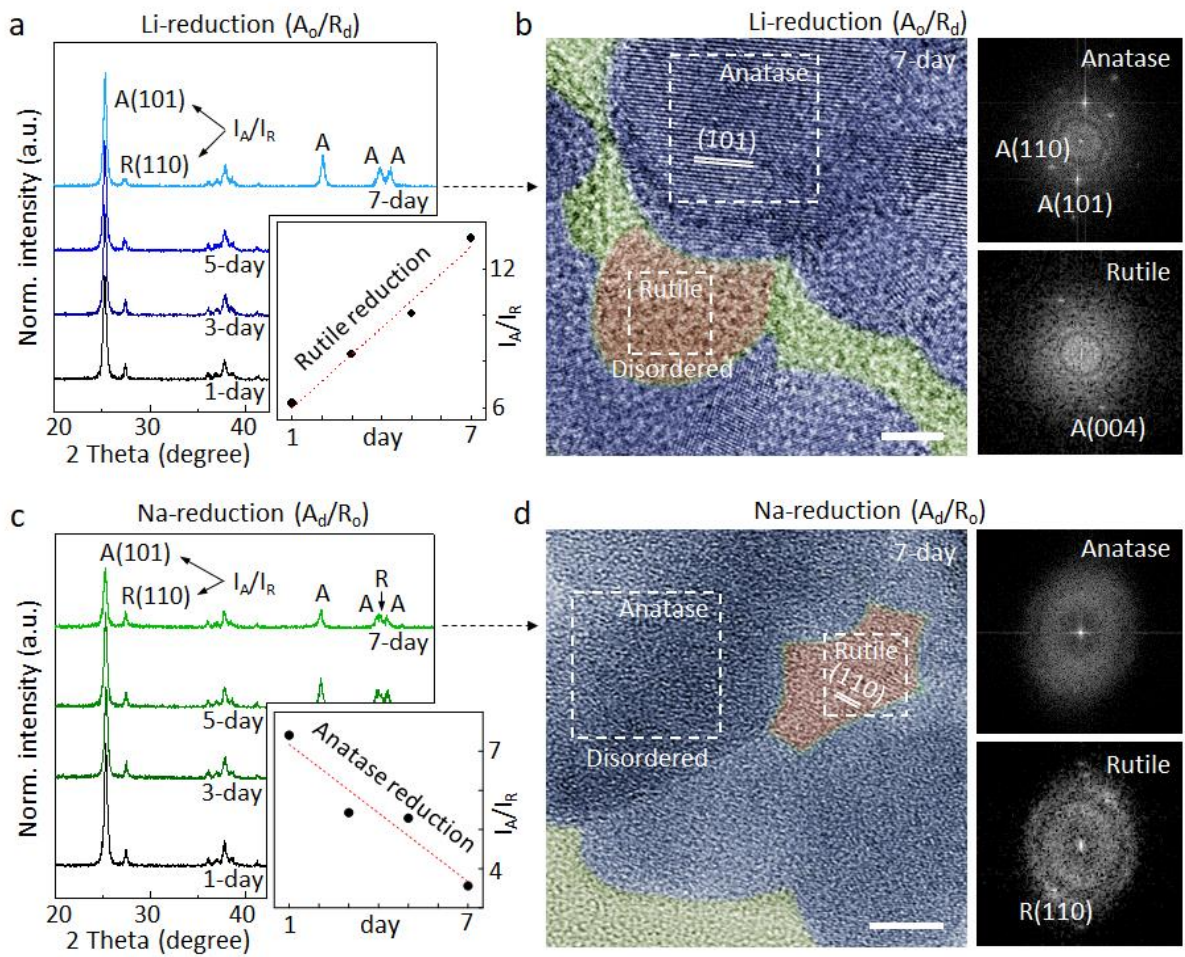

**Fig. 2.** Crystal and chemical structural characterization of alkali metal-treated phase-mixed TiO$_2$. XRD patterns of (a) 7-day Li-reduction (A$_o$/R$_d$) and (c) 7-day Na-reduction (A$_d$/R$_o$). Inset: Variation of I$_A$/I$_R$ values corresponding to the peak intensity ratio of anatase (101) to rutile (110). HRTEM images of (b) 7-day Li-reduction (A$_o$/R$_d$) and (d) 7-day Na-reduction (A$_d$/R$_o$). Scale bars are 5 nm. Inset: Electron-diffraction FFT patterns of anatase and rutile phases.

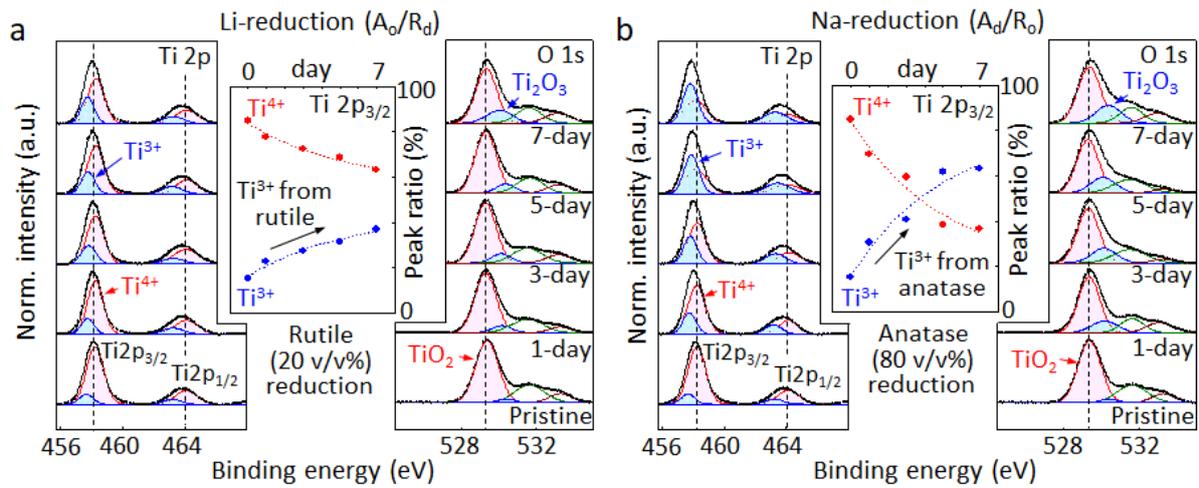

**Figure 3.** XPS core-level spectra of Ti 2p and O 1s from (a) Li-reduced ($A_o/R_d$) and (b) Na-reduced ($A_d/R_o$) phase-mixed $TiO_2$. Inset: Deconvoluted peak ratio of $Ti^{4+}$ and $Ti^{3+}$ from the Ti $2p_{3/2}$ peak with Li- and Na-reduction time evolution.

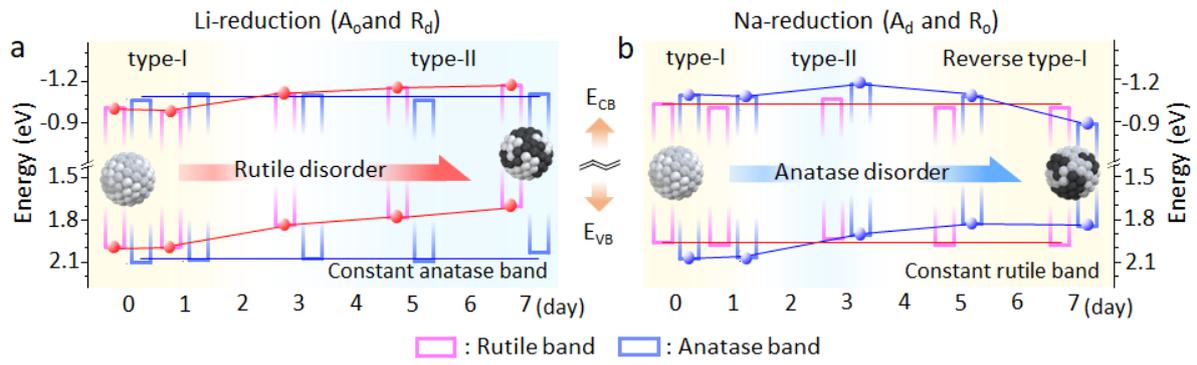

**Figure 4.** Restructured energy band positions of phase-mixed TiO$_2$ with the reduction time evolution. (a) Heterojunction band structures of ordered anatase and disordered rutile TiO$_2$ interface with Li-reduction. (b) Heterojunction band structures of disordered anatase and ordered rutile TiO$_2$ interface with Na-reduction.

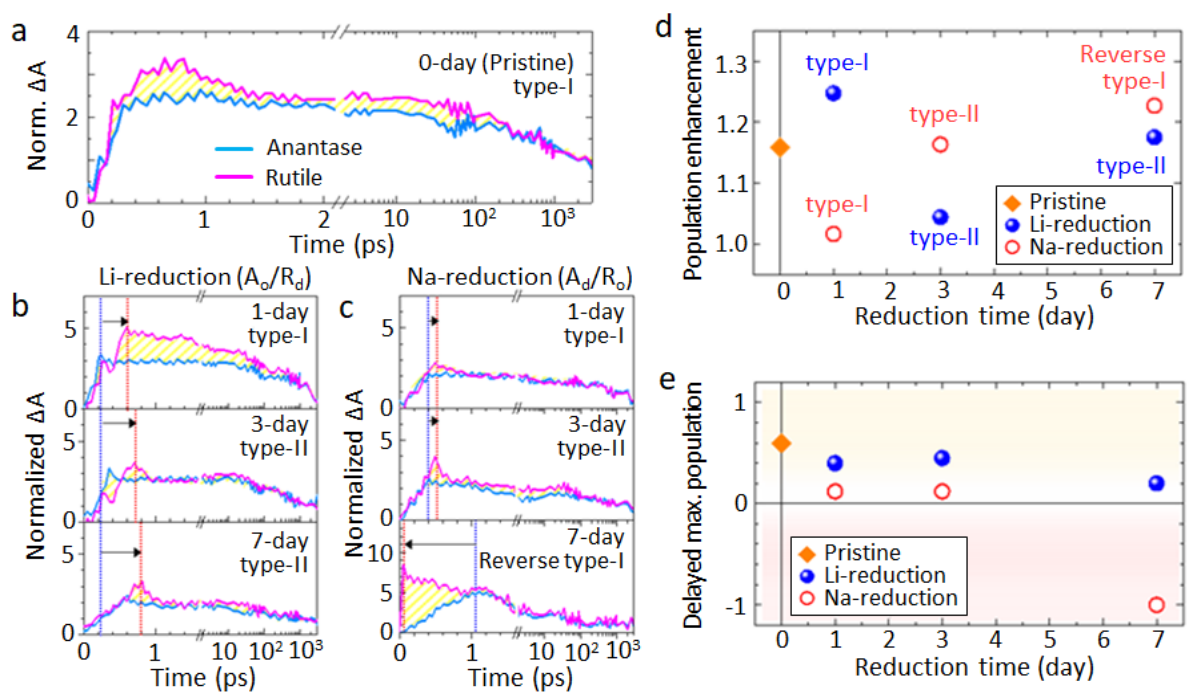

**Figure 5.** Femtosecond transient absorption spectra of phase-mixed $TiO_2$. Normalized femtosecond transient absorption spectra of (a) pristine phase-mixed $TiO_2$, (b) Li-reduced ($A_o/R_d$) and (c) Na-reduced ($A_d/R_o$) phase-mixed $TiO_2$. The 300 nm pump photon excitation energy applied. (d) Population enhancement tendency. (e) Delayed maximum population.

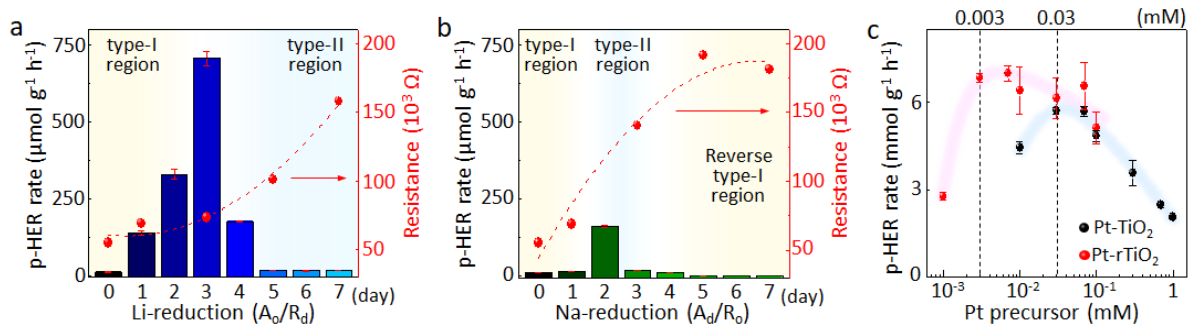

**Figure 6.** Photocatalytic hydrogen evolution reaction (p-HER). The p-HER rate and equivalent serial resistance of (a) Li-reduced ($A_o/R_d$) and (b) Na-reduced ($A_d/R_o$) phase-mixed $TiO_2$. (c) The p-HER rates of Pt-$TiO_2$ and Pt-r$TiO_2$ with a 10 times smaller amount of Pt co-catalyst.

25